\documentclass[11pt]{amsart}
\usepackage{a4wide}

\def\endproof{\hfill$\Box$}

\def\M{{\mathcal M}}

\def\K{{\mathcal K}}
\newcommand{\DD}{{\mathcal D}}

\newtheorem{theorem}{Theorem}
\newtheorem{prop}[theorem]{Proposition}
\newtheorem{lemma}[theorem]{Lemma}
\newtheorem{corollary}[theorem]{Corollary}
\newtheorem{example}[theorem]{Example}

\newcommand{\KK}{{\mathcal K}}
\newcommand{\al}{\alpha}
\newcommand{\ga}{\gamma}

\newcommand{\pal}{\partial}

\begin{document}

\title[Simple Elliptic Singularities: a note on their $G$-function]
{Simple Elliptic Singularities: a note on their $G$-function}

\author{Ian A.B. Strachan}
\date{23$^{\rm rd}$ September 2010: updated 15$^{\rm th}$ September 2011}
\address{School of Mathematics and Statistics \\ University of Glasgow\\
Glasgow G12 8QQ\\ U.K.}

\email{ian.strachan@glasgow.ac.uk}


\begin{abstract}
The link between Frobenius manifolds and singularity theory is well known, with the simplest examples
coming from the simple hypersurface singularities. Associated with any such manifold is a function known
as the $G$-function. This plays a role in the construction of higher-genus terms in various theories.
For the simple singularities the $G$-function is known explicitly: $G=0\,.$ The next class of singularities,
the unimodal hypersurface or elliptic hypersurface singularities consists of three examples,
$\widetilde{E}_6\,,\widetilde{E}_7\,,\widetilde{E}_8$ (or equivalently $P_8\,,X_9\,,J_{10}$).
Using a result of Noumi and Yamada on the flat structure on the space of versal deformations of these singularities the
$G$-function is explicitly constructed for these three examples. The main property is that the function depends on
only one variable, the marginal (dimensionless) deformation variable. Other examples are given based on the foldings of
known Frobenius manifolds. Properties of the $G$-function under the
action of the modular group is studied, and applications within the theory of integrable systems are discussed.
\end{abstract}

\maketitle

\tableofcontents

\bigskip

\section{Introduction}

The existence of a Frobenius manifold structure on the space of versal deformations of simple isolated
singularities ties together many apparently different areas of mathematics and mathematical physics.
K. Saito proved the existence of a flat-structure on such spaces \cite{saito}, and it was later realized that this
work provides one of the three main constructions of Frobenius manifolds. Such geometric structures then
provide the link with the original ideas of Topological Quantum Field Theory (TQFT) and the theory of
integrable systems (though such links were also implicit in the work of Noumi via the structure of the
associated Gauss-Manin equations \cite{Noumi1}).

Given a semi-simple Frobenius manifold there exists an intriguing function on it known as the $G$-function.
Physically this plays a role in
the construction of genus one objects from genus zero data. Thus
in TQFT it appears in (and is in fact defined by) the genus one contribution to the free
energy of the field theory \cite{DW}, i.e.
\begin{equation}
\mathcal{F}_1({\bf t},{\bf t}_X) = \frac{1}{24} \log\det \left(c_{\alpha\beta\gamma}({\bf t}) t^\gamma_X\right) + G({\bf t})\,;
\label{F1}
\end{equation}
in enumerative geometry it governs
genus one Gromov-Witten invariants; in integrable systems it
appears in the first order deformation of bi-Hamiltonian
structures. Within singularity theory is plays a role in the next-to-leading order
expansions of oscillatory integrals. The aim of this note is to give an explicit
description of this $G$-function for the elliptic isolated singularities (or parabolic unimodal singularities), ${\widetilde E}_6\,, {\widetilde E}_7\,$ and
${\widetilde E}_8\,$ (or $P_8\,,X_9\,,J_{10})$) in terms of the flat structures on the base space of the unfolding spaces of these singularities.

For the simple isolated singularities $A_n\,,D_n$ and $E_{6,7,8}$ the answer was found by Givental \cite{Gi1}: $G=0\,.$
A unified proof of this result, together with the corresponding formulae for boundary hypersurface singularities
was given by the author in \cite{me}\,. This utilized the description of these spaces as Coxeter group
orbit spaces $\mathbb{C}/W$ for $W$ an irreducible finite Coxeter group. The proof of this result will be repeated here
both for completeness and because it provides a simpler model for the main theorem that will be proved for the elliptic singularities.

\section{The $G$-function}

It will be assumed that the reader is familiar with the definition of a Frobenius manifold: the details may be found in
\cite{D1}\,. The governing equations for the function $G$ appearing in (\ref{F1}) were obtained by Getzler \cite{Ge} and are the following overdetermined
set of linear equations \cite{DZ2}:

\begin{equation}
\sum_{1\le \al_1,\al_2,\al_3,\al_4\le n} z_{\al_1} z_{\al_2}
z_{\al_3} z_{\al_4} \Delta_{\al_1\al_2\al_3\al_4}=0
\label{Getz}
\end{equation}
where
\begin{eqnarray*}
\Delta_{\al_1\al_2\al_3\al_4}&=&
3\,c^{\mu}_{\al_1\al_2}\,c^{\nu}_{\al_3\al_4}
\,\frac{\pal^2 G}{\pal t^\mu\pal t^{\nu}}-4\,
c^{\mu}_{\al_1\al_2}\,c^{\nu}_{\al_3\mu}
\,\frac{\pal^2 G}{\pal t^{\al_4}\pal t^{\nu}}
-c^{\mu}_{\al_1\al_2}\,c^{\nu}_{\al_3\al_4\mu}
\,\frac{\pal G}{\pal t^{\nu}}+
\nonumber\\
&& 2\,c^{\mu}_{\al_1\al_2\al_3}\,c^{\nu}_{\al_4\mu} \,\frac{\pal
G}{\pal t^{\nu}}+\frac16
c^{\mu}_{\al_1\al_2\al_3}\,c^{\nu}_{\al_4\mu\nu} +\frac1{24}
c^{\mu}_{\al_1\al_2\al_3\al_4}\,c^{\nu}_{\mu\nu}- \frac14
c^{\mu}_{\al_1\al_2\nu}\,c^{\nu}_{\al_3\al_4\mu}\,.
\end{eqnarray*}
In \cite{DZ2} Dubrovin and Zhang (following from conjectures of Givental \cite{Gi1}) proved
that for semisimple Frobenius manifolds the $G$-function is given by the formula
\begin{equation}
G=\log  \frac{ \tau_I}{J^{1/24}}
\label{Gsolution}
\end{equation}
where $\tau_I$ is the isomonodromic $\tau$-function and $J$ is the Jacobian of the
transformation between canonical and flat-coordinates. Furthermore, it was shown that \cite{DZ2}:
\begin{theorem}
For an arbitrary semisimple Frobenius manifold the system (\ref{Getz}) has a
unique, up to an additive constant, solution $G=G(t^2, \dots, t^n)$
satisfying the quasihomogeneity condition
\begin{equation}
\label{anomaly}
{\mathcal L}_E\,G=\ga
\end{equation}
with a constant $\ga$. This solution is given by the formula (\ref{Gsolution})
where $\tau_I$ is the isomonodromic tau-function and
\[
J=\det \left(\frac{\partial t^\alpha}{\partial u^i} \right)
\]
is the Jacobian of the transform from the canonical coordinates to
the flat ones. The scaling anomaly $\ga$ in (\ref{anomaly}) is given by
the formula
\[
\ga =-\frac{1}{4} \sum_{\alpha=1}^n \mu_\alpha^2 +\frac{n\, d}{48}
\]
where
\[
\mu_\alpha = q_\alpha -\frac{d}{2}, ~\alpha=1, \dots, n.
\]
\end{theorem}
\noindent A simple extension of this result appeared in \cite{DZ3}:

\begin{theorem}\label{virasoro}
The derivatives of the $G$-function along the
powers
of the Euler vector field are given by the following formulae
\begin{eqnarray*}
&&\mathcal{L}_e G=0,
\\
&&
\mathcal{L}_E G= \frac{n\, d}{48} -\frac{1}{4} {\rm tr}\, \mu^2,
\\
&&
\mathcal{L}_{E^k}G =-\frac{1}{4}{\rm tr}\,\left(\mu \, (\mu\, {\mathcal U}^{k-1}
+{\mathcal U}\mu\, {\mathcal U}^{k-2} + \dots + {\mathcal U}^{k-1}\mu)\right)
\nonumber\\
&&\quad
-\frac{1}{24}\left< (\mu\, {\mathcal U}^{k-2}+ {\mathcal U}\mu\, {\mathcal U}^{k-3}+
\dots + {\mathcal U}^{k-2}\mu)\,E -\frac{d}{2}\, {\mathcal U}^{k-2}\,E, H\right>,
\\
&&\quad  k\geq 2\nonumber
\end{eqnarray*}
where
\[
\mu = diag\{{\mu_\alpha}\}\,,\qquad {\mathcal U}^\nu_\mu=c^\nu_{\mu \sigma} E^\sigma\,,\qquad H=c_\nu^{\nu\alpha}\partial_\alpha
\]
and $\langle X,Y \rangle = \eta_{\alpha\beta} X^\alpha Y^\beta\,.$
\end{theorem}
It is this last theorem that provides a practical way to construct the $G$-function: the first
$N$-equations give a simple linear system for the $N$-first derivatives. Solving these linear
equations gives the first derivatives of the $G$ function from which the $G$ function itself may be found by integration.

\bigskip

\section{Elliptic Singularities}

The simple elliptic singularities have been studied from a number of different perspectives:
early results were proved by Arnold \cite{Arnold}, K. Saito \cite{saito2} and Looijenga \cite{Loo} in the early 1970's, with the
properties of the associated Gauss-Manin system being studied by Noumi in the 1980's \cite{Noumi1}. In the
late 1980's/early 1990's they were studied from the point of view of topological Landau-Ginzburg theories,
and hence as specific examples of Frobenius manifolds \cite{VW,KTS}. Different communities refer to the same object in different
ways: the canonical form of an isolated singularity, together with its versal deformation is called the Landau-Ginzburg
superpotential, and the Jacobi ring is called the chiral ring.

The canonical forms of the simple elliptic hypersurface singularities are
(ignoring quadratic terms which will play no
role in what follows):
\[
\begin{array}{lrcl}
{\widetilde E}_6: & f({\bf z}) & = & z_1^3 + z_2^3 + z_3^3\,,\\
{\widetilde E}_7: & f({\bf z}) & = & z_1^4 + z_2^4\,,\\
{\widetilde E}_6: & f({\bf z}) & = & z_1^6 + z_2^3 + z_3^3\,.
\end{array}
\]
Versal deformation of these take the form
\[
W({\bf z},{\bf s}) = f({\bf z}) + \sum_{\nu \in I} s_\nu {\bf z}^\nu
\]
(using a multi-index notation so ${\bf z}^\nu = z_1^{\nu_1} \ldots z_n^{\nu_n}$ and some indexing set $I$).
These are then used to define the Jacobi ring
\[
\mathcal{R}\cong \mathbb{C}[z_1\,,\ldots\,, z_n] / (\partial_{z_1} W\,, \ldots \,, \partial_{z_n} W)
\]
and it follows from the Saito construction that the base space of the universal unfolding carries a structure
of a Frobenius manifold, and so, in particular, there exists a naturally defined flat metric (the so-called
Saito metric) on this manifold.
The construction of this flat metric and associated flat coordinates is subtle: it relies on K. Saito's
notion of a primitive form. For the simple singularities the flat coordinates are polynomial functions of
the deformation variable $\{ {\bf s}^\nu\in I\}\,,$ but for the simple elliptic singularities the relationship
is far more complicated \cite{NY}.

Explicitly, the superpotential are\footnote{For the ${{\bf s}}$-variables {\sl only} $s_i=s^i\,,$ this being for notational convenience.}
\begin{equation}
\begin{array}{lrcl}
{\widetilde E}_6: & W({\bf z},{\bf s}) & = & z_1^3 + z_2^3 + z_3^3 + s_8 z_1 z_2 z_3 \\
&&&+ s_7 z_1 z_2 + s_6 z_1 z_3 + s_5 z_2 z_3 + s_4 z_1 + s_3 z_2 + s_2 z_3 + s_1\,,\\ \\
{\widetilde E}_7: & W({\bf z},{\bf s}) & = & z_1^4 + z_2^4 + s_9 z_1^2 z_2^2 \\
&&&+ s_8 z_1^2 z_2 + s_7 z_1 z_2^2+ s_6 z_1^2 + s_5 z_1 z_2 + s_4 z_2^2 + s_3 z_1 + s_2 z_2 + s_1\,,\\ \\
{\widetilde E}_8: & W({\bf z},{\bf s}) & = & z_1^6 + z_2^3 + z_3^3 + s_{10} z_1^4 z_2 \\
&&& +s_9 z_1^3 z_2 + s_8 z_1^4 + s_7 z_1^2 z_2 + s_6 z_1^3 + s_5 z_1 z_2 + s_4 z_1^2 + s_3 z_2 + s_2 z_1 + s_1\,,
\end{array}
\label{sdefn}
\end{equation}
and hence the associated Frobenius manifolds are of dimension $8\,,9$ and $10$ (which is why these singularities
are also named $P_8\,, X_9\,, J_{10}$ respectively).
Assigning weights to the variables in such a way that $W$ is of weight 1 gives the following scaling
dimensions to the deformation variables:
\begin{equation}
\begin{array}{lcl}
{\widetilde E}_6 & : & \left( 1\,, \frac{2}{3}\,,\frac{2}{3}\,,\frac{2}{3}\,,\frac{1}{3}\,,\frac{1}{3}\,,\frac{1}{3}\,,0\right),\\ &&\\
{\widetilde E}_7 & : & \left( 1\,, \frac{3}{4}\,,\frac{3}{4}\,,\frac{1}{2}\,,\frac{1}{2}\,,\frac{1}{2}\,,\frac{1}{4}\,,\frac{1}{4}\,,0\right),\\ &&\\
{\widetilde E}_8 & : & \left( 1\,, \frac{5}{6}\,,\frac{2}{3}\,,\frac{2}{3}\,,\frac{1}{2}\,,\frac{1}{2}\,,\frac{1}{3}\,,\frac{1}{3}\,,\frac{1}{6}\,,0\right)\,.\\
\end{array}
\label{data}
\end{equation}
Note (in contrast to simple singularities) that there is, in each case, a distinguished deformation whose scaling degree is zero.
This is a so-called marginal deformation variable. This
marginal variable will be denoted $s$ (rather than $s^n$) in what follows.
Note, for the ring structure to be finite dimensional one requires:
\[
\begin{array}{lrl}
{\widetilde E}_6: &\displaystyle{ 1+ \frac{1}{27} s^3 }&\neq 0 \,,\\ &&\\
{\widetilde E}_7: &\displaystyle{ 1- \frac{1}{4} s^2 }&\neq 0 \,,\\ && \\
{\widetilde E}_8: &\displaystyle{ 1+ \frac{4}{27} s^3 }&\neq 0 \,
\end{array}
\]
and it will be assumed for the rest of this paper that these inequalities hold.

The flat coordinates $\{ t^i \}$ for these singularities have been calculated by Noumi and Yamada \cite{NY}. These flat coordinates are transcendental
in the variable $s$ and polynomial in the remaining variables. The complete set may be found in \cite{NY}, but here only the relationship
between $t$ and $s$ will be given:
\begin{equation}
\begin{array}{lrcl}
{\widetilde E}_6\,: & t  & =& s \,\displaystyle{\frac{ {}_2F_1(\frac{2}{3},\frac{2}{3},\frac{4}{3},-\frac{1}{27} s^3 )}
{ {}_2F_1(\frac{1}{3},\frac{1}{3},\frac{2}{3},-\frac{1}{27} s^3 )}}\,,\\
&&&\\
{\widetilde E}_7 \,: & t  & =& s \,\displaystyle{\frac{ {}_2F_1(\frac{3}{4},\frac{3}{4},\frac{3}{2},\frac{1}{4} s^2 )}
{ {}_2F_1(\frac{1}{4},\frac{1}{4},\frac{1}{2},\frac{1}{4} s^2 )}}\,,\\
&&&\\
{\widetilde E}_8\,: & t  & =& s \,\displaystyle{\frac{ {}_2F_1(\frac{5}{12},\frac{11}{12},\frac{4}{3},-\frac{4}{27} s^3 )}
{ {}_2F_1(\frac{1}{12},\frac{7}{12},\frac{2}{3},-\frac{4}{27} s^3 )}}
\end{array}
\label{schwartz}
\end{equation}
where ${}_2F_{1}$ is the hypergeometric function. The inverse functions are given by the Schwartz triangle functions.

The Frobenius potential has been explicitly calculated by several authors \cite{VW,KTS,NY} for ${\widetilde{E}}_6\,,$ and by \cite{KTS}
for ${\widetilde{E}}_7\,.$ The potential for ${\widetilde E}_8$ does not appear to have been calculated explicitly. Given that
the ring structure and flat coordinates are already known it would just be a computational exercise, albeit a very complicated exercise to calculate it:
one can easily construct the multiplication table (i.e. the $(2,1)$-tensor $c_{ij}^k({\bf s})$) in terms of the $s^a$-variables defined in equation (\ref{sdefn})
and then use the explicit transformation between the $s^a$-coordinates and the flat coordinates $t^\mu$ found in \cite{NY} to find this tensor in terms of the
flat coordinates from whence one could integrate three times to obtain the potential function.
Thus is all cases one could, in principle, calculate the $G$-function explicitly.
It will turn out, however, that using the concept of an $F$-manifold one can bypass such a computational procedure. This approach
relies on the study of the singular loci within the manifold where the Frobenius multiplication ceases to be semi-simple.

\section{Multiplication on caustics and the $G$-function}

By definition, a massive Frobenius manifold $\M$ has a semisimple multiplication
on the tangent space at {\sl generic} points of $\M\,.$ The set of points where the multiplication
is not semisimple is known as the caustic, and will be denoted $\K\,.$ This is an
analytic hypersurface in $\M\,,$ which may consist of a number of
components (possibly highly singular),
\[
\K = \bigcup_{i=1}^{\#\K_i} \K_i\,.
\]
The set of smooth points in $\K$ will be denoted $\K_{reg}\,.$

The simplest case is where the multiplication on the caustic $\K_i$ is
of the type $A_1^{n-2} \,I_2(N_i)\,,$ i.e. the multiplication decomposes into
$n-2$ one-dimensional algebras and a single two-dimensional algebra
based on the Coxeter group $I_2(N)\,.$ The following theorem
studies the behaviour of the $G$-function near such caustics.

\begin{theorem}\label{ClausThm}\cite{He}
Let $(M,\circ,e,E,g)$ be a simply connected massive Frobenius manifold.
Suppose that at generic points of the caustic $\KK_i$ the germ of the
underlying F-manifold is of type $I_2(N_i)A_1^{n-2}$ for one
fixed number $N_i\geq 3$.

a) The form $d\, \log \tau_I$ has a logarithmic pole along $\KK_i$ with
residue $-\frac{(N_i-2)^2}{16 N_i}$ along $\K_i\cap\KK_{reg}$.

b) The form $-\frac{1}{24} d \log J$ has a logarithmic pole along $\KK_i$
with residue $\frac{N_i-2}{48}$ along $\K_i\cap\KK_{reg}$.

c) The G-function extends holomorphically over $\KK_i$ iff $N_i=3$.
\end{theorem}

\noindent Not all caustics within Frobenius manifolds are of this type, for example
the Frobenius manifold describing the quantum cohomology of $\mathbb{P}^1$ is not
of this type. This has a logarithmic caustic and an analogous result on the behaviour
of the $G$-function near such a caustic may be proved \cite{me}.

The above theorem is a local result - it gives the behaviour of the $G$-function near
a particular type of caustic. However, for certain classes of Frobenius manifold, one
may \lq glue\rq~such local behaviour together to determine the $G$-function globally purely from the
knowledge of such local behaviour.

\section{The $G$-function for Coxeter groups}

The Saito construction \cite{saito} of a Frobenius manifold structure on the orbit space $ \mathbb{C}^n/W$
(where $W$ is a Coxeter group) is given in \cite{D1}. The only parts of that construction that
will be required in this section are the following:

\begin{itemize}
\item{}
in flat-coordinates, the prepotential, and hence
the structure functions of the Frobenius algebra, are polynomial functions;
\item{}
the Euler vector field takes the form
\[
E=\frac{1}{h} \sum_{\alpha=1}^n d_\alpha t^\alpha \frac{\partial~}{\partial
t^\alpha}\,\qquad\qquad d_\alpha>0
\]
where the $d_\alpha$ are the exponents of the Coxeter group and $h$ is the Coxeter number of $W\,.$
\end{itemize}

\noindent The components of the caustic $\KK$ for such orbit spaces are given in terms of
quasihomogeneous polynomials $\kappa_i$ such that $\kappa_i^{-1}(0)=\K_i\,.$ The $F$-manifold
structure on these caustics is known, the multiplication is of type $I_2(N_i) A_1^{n-2}\,,$
and this enables Theorem 3 to be used. The data $N_i$ is given in Table 1.
It can be extracted with some work from \cite{H2} Theorem 5.22, which builds on
\cite{Gi2}.

\begin{prop}\label{GCoxeter} The $G$-function on $ \mathbb{C}^n/W$ takes the form
\[
G=-\frac{1}{24} \frac{(N_1-2)(N_1-3)}{N_1} \log \kappa_1
\]
and the constant $N_1\,,$ which depends on the Coxeter group
$W\,,$ is given in table 1.
\end{prop}

\bigskip

\begin{table}
\begin{center}
\begin{tabular}{c|c|c}
Coxeter Group $W$ & Number of caustics & Values of $N_i$ \\ \hline
$A_n\,,D_n\,,E_{6,7,8}$ & $1$ & $N_1=3$ \\
& & \\
$B_n$ & $2$ & $N_1=4\,,N_2=3$\\
& & \\
$F_4$ & $3$ & $N_1=4\,,N_2=N_3=3$\\
& & \\
$H_3$ & $2$ & $N_1=5\,,N_2=3$\\
& & \\
$H_4$ & $2$ & $N_1=5\,,N_2=3$\\
& & \\
$I_2(h)$ & $1$ & $N_1=k$\\
\end{tabular}
\end{center}
\vskip 5mm
\caption{Data on the caustics of the Frobenius manifold $\mathbb{C}^n/W$}
\end{table}

\noindent{\bf Proof~} From Theorem \ref{virasoro} it follows that the only singularities of $dG$
are on caustics, and it is known that the multiplication on the caustics
of a Coxeter group orbit space Frobenius manifold is of the form where Theorem \ref{ClausThm}
may be applied (see \cite{Gi2} and \cite{H2} Theorem 5.22).

It follows from the polynomial nature of the structure functions
and Theorem 2 that all first derivatives $\partial G/\partial t^\alpha$ are rational
functions. Hence, on integrating, $G$ takes the schematic form
\[
G(t)={\rm rational~function~} + {\rm logarithmic~singularities}\,.
\]
By Theorem 3, the only singularities that $G$ has are logarithmic singularities on $\KK_i\,.$
Thus the rational functions must be polynomial. However, the only polynomial
function compatible with the symmetry (\ref{anomaly}) is a constant (this uses the fact that
the exponents of the Coxeter group are all positive). Since $G$ is only defined up
to a constant anyway one has:
\[
G(t) = -\frac{1}{24}\sum_{i=1}^{\#\KK_i} \frac{(N_i-2)(N_i-3)}{N_i}\log\kappa_i
\]
and hence
\[
\gamma = -\frac{1}{24}\sum_{i=1}^{\#\KK_i} \frac{(N_i-2)(N_i-3)}{N_i}E\left(\log\kappa_i\right)\,.
\]
Using the data in table 1,
and in particular the fact that in all cases
there is at most one caustic, denoted $\K_1\,,$ with $N_i>3\,,$ the result follows.
\endproof

\bigskip

\begin{corollary}
For the Coxeter groups $A_n\,,D_n\,,E_{6,7,8}$ and hence for the simple hypersurface singularities,
\[
G({\bf t})=0\,.
\]
\end{corollary}
\noindent The first proof of this result, by a different method, was given by Givental \cite{Gi1}.

\section{The $G$-function for the simple Elliptic Singularities}

The following facts on the Frobenius manifold structure of the elliptic singularities will be used:
\begin{itemize}
\item{}
in flat-coordinates, the prepotential $F$, and hence
the structure functions of the Frobenius algebra, are polynomial functions in the variable $t^\alpha\,,\alpha=1\,,\ldots\,,n-1\,.$
\item{}
the Euler vector field takes the form
\[
E=\sum_{\alpha=1}^n d_\alpha t^\alpha \frac{\partial~}{\partial
t^\alpha}\,,
\]
where the $d_\alpha\,(=1-q_\alpha)$ are given by equation (\ref{data});
\item{}
$\mathcal{L}_E F = 2 F$ modulo quadratic terms (equivalently, $d=1\,).$
\end{itemize}
Note, it immediately follows that $\ga=0$ for these singularities.

\begin{lemma}
For the simple elliptic singularities ${\widetilde{E}}_{6,7,8}\,,$
\[
G=G(t)\,,
\]
i.e. it is a function only of the marginal deformation variable.
\end{lemma}

\noindent{\bf Proof~} One may adopt the proof of Proposition \ref{GCoxeter} to apply to the non-marginal
deformation variables using the above facts. Thus
\[
G({\bf t})={\rm rational~function~} + {\rm logarithmic~singularities}
\]
as functions of the non-marginal variables. However, it is known that for hypersurface singularities
that $N_i=3$ and hence Theorem \ref{ClausThm} may be applied \cite{H2}. This eliminates the possibility of $G$ being
a rational function or having a logarithmic singularity. Thus all that is left is the
possibility that $G=G(t)\,.$
\endproof

\medskip

Note that this is consistent with the result $\mathcal{L}_E G=0$ since the marginal variable
is dimensionless. To find $G$ one only has to consider the Virasoro constraint equation $\mathcal{L}_{E^2} G$ to
construct an ordinary differential equation for the function $G$ itself.

\begin{prop}\label{dGelliptic} For each $\mu\in\{2\,,\ldots\,,n-1\}\,,$ the function $G$ satisfies the ordinary
differential equation
\[
\frac{d G}{dt} = \frac{1}{2 d^\mu d^{\mu^\star}} \sum_{\mu=1}^{n}
\left[ \frac{1}{24} \left\{ \left( d^{\mu^{\phantom{\star}}}\right)^2 + \left( d^{\mu^\star}\right)^2 \right\}
- \frac{1}{2} \left\{ \frac{1}{2} - d^\nu\right\}^2
\right]\, c^\nu_{\nu\mu\,,{\mu^\star}}
\]
where $\mu^\star=n+1-\mu$ for each index $\mu\,.$
\end{prop}

\medskip

\noindent{\bf Proof~} From Theorem \ref{virasoro}:
\[
\mathcal{L}_{E^2}G =-\frac{1}{4}{\rm tr}\,\left[\mu \, \left(\mu\, {\mathcal U}
+{\mathcal U}\mu\right)\right]
-\frac{1}{24}\left\langle \mu \,E -\frac{1}{2}\,E, H\right\rangle\,.
\]
Unpacking the various definitions and using the above lemma results in the formula
\[
\left< E,E \right> \frac{dG}{dt} = \sum_{\nu,\mu} d^\mu t^\mu c^\nu_{\nu\mu} \left[\frac{1}{24}d^\mu - \frac{1}{2} \left(\frac{1}{2} - d^\nu\right)^2 \right]\,.
\]
Since $d^n=0$ the term $\left< E,E \right>$ is independent of $t$ and hence the $t$ behaviour factors out in the left hand side (note, it is certainly not
a priori obvious that such a factorization should occur in the right hand side of the above formula). Since $\left< E,E \right>$ is quadratic
in the non-marginal variables one only has to pick out the quadratic terms on the right hand side. However there is already an explicit linear
term, so one just requires the linear term in the expansion of the structure functions around the point ${\bf 0}^\star=(0\,,\ldots\,,0\,,t)\,.$
Any higher-order terms must eventually cancel.

Note that
\[
\left.c^\nu_{\nu\mu,\mu^\star}\right|_{{\bf 0}^\star} = c^\nu_{\nu\mu,\mu^\star}
\]
since the right hand side is dimensionless and hence cannot depend on the non-marginal variables. Thus picking out the coefficient of $t^\mu t^{\mu^\star}$ gives the final result.

Note that different values of $\mu$ must result in the same differential equation for $G\,.$

\endproof

By summing over $\mu$ one can obtain a more symmetric differential equation for $G\,,$ namely
\[
2 \left( \sum_\mu d^\mu d^{\mu^\star} \right) \frac{dG}{dt} =
\sum_\mu \left[ -\frac{1}{12} + \frac{5}{12} d^\mu d^{\mu^\star} \right] \sum_\nu c_{\nu\nu^\star\mu,\mu^\star}\,.
\]
However the equation derived in Proposition \ref{dGelliptic} is more useful from a computational point of view as it requires knowledge of fewer
structure functions. The $G$ function can now be written down without further calculation for the ${\widetilde{E}}_{6,7}$
manifolds, since the corresponding prepotential is explicitly known. All one has to do is to pick the appropriate terms and substitute them
into the above formula. However, a direct approach will be given below. This has the advantage that it will work for
all of the simple elliptic singularities, including $\widetilde{E}_8$ for which the full prepotential is unknown.

\begin{theorem}
The $G$-function for the elliptic singularities $\widetilde{E}_{6,7,8}$ are:

\begin{itemize}

\item[$\widetilde{E}_6\,:$]   $G(t)=\displaystyle{-\frac{1}{24} \log \left\{ \frac{    \left[ s(t)^\prime \right]^2 }{1+\frac{1}{27} s(t)^3 }\right\}}\,;$

    \vskip 5mm

\item[$\widetilde{E}_7\,:$]   $G(t)=\displaystyle{-\frac{1}{24} \log \left\{ \frac{    \left[ s(t)^\prime \right]^\frac{3}{2} }{1-\frac{1}{4} s(t)^2 }\right\}}\,;$

\vskip 5mm

\item[$\widetilde{E}_8\,:$]   $G(t)=\displaystyle{-\frac{1}{24} \log \left\{ \frac{    s(t)^\prime  }{\left[1+\frac{4}{27} s(t)^3\right]^\frac{5}{6} }\right\}}\,.$
\end{itemize}
In each case, $s(t)$ is the inverse of the functions given by equation (\ref{schwartz}).
\end{theorem}

\noindent{\bf Proof} Full details will only be given here for $\widetilde{E}_6\,;$ the slight modifications which occur for
${\widetilde{E}}_{7,8}$ will be outlined at the end.

On using Proposition \ref{dGelliptic} with $\mu=2$ (so $\mu^\star=7$) one obtains
\begin{equation}
\frac{dG}{dt} = \frac{1}{48} \sum_{\nu=1}^n c_{\nu\nu^\star 2,7}(t)\,.
\label{GE6}
\end{equation}
The calculation of the $c_{\nu\nu^\star\mu,\mu^\star}$ is straightforward. This is first done in the $\{{\bf s}\}$-variables defined by equation (\ref{sdefn}) and then
converted, using a changes of variable, to the $\{ {\bf t} \}$-variables calculated in \cite{NY}. Thus
\[
c_{\nu\nu^\star\mu,\mu^\star}(t) = \frac{\partial~}{\partial t^{\mu^\star}} \left[
\frac{\partial s^a}{\partial t^\mu}
\frac{\partial s^b}{\partial t^\nu}
\frac{\partial s^c}{\partial t^{\nu^\star}}
c_{abc}({\bf s})\right]\,.
\]
At this stage one requires the explicit forms of the flat coordinates. For $\widetilde{E}_6$ these are
\[
t=s \,\displaystyle{\frac{ {}_2F_1(\frac{2}{3},\frac{2}{3},\frac{4}{3},-\frac{1}{27} s^3 )}
{ g(u) }}
\]
where $g(u)={}_2F_1(\frac{1}{3},\frac{1}{3},\frac{2}{3},u)$ and $u=-\frac{1}{27} s^3\,,$ together with
\begin{eqnarray*}
s^a & = & (1-u)^\frac{2}{3} g(u) t^a\,, \phantom{+ O(t^2)} \qquad a=5\,,6\,,7\,,\\
s^a & = & (1-u)^\frac{1}{3} g(u) t^a + O(t^2)\,,\qquad a=2\,,3\,,4\,, \\
s^1 & = & t^1 + \frac{1}{9} s^2 (1-u) g(u) g^\prime(u) \left[t^2 t^7 + t^3 t^6 + t^4 t^5 \right]+O(t^3)\,.
\end{eqnarray*}
Here the $O(t^2)$ and $O(t^3)$ terms are quadratic and cubic in the non-marginal variables. One may also
write these in terms of Jacobi forms \cite{Satake}.

Another computational simplification now occurs: the term $c_{\nu\nu^\star\mu,\mu^\star}$ is dimensionless
(its scaling dimension is $3-d-d^\mu-d^{\mu^\star}-d^\nu-d^{\nu^\star}$ and since $d=1$ for these models,
and $d^\mu+d^{\mu^\star}=1\,,$ this term is dimensionless and hence a function of $t$ alone). Thus the right
hand side may be evaluated at the point ${\bf 0^\star}$ without affecting the calculation (higher-order terms must eventually
cancel). So, for example,
\[
\left.
\frac{\partial s^a}{\partial t^\mu}\right|_{\bf 0^\star} = \left\{
\begin{array}{lcl}
\displaystyle{(1-u)^\frac{2}{3} g(u) \delta^a_\mu }& \qquad & a=5\,,6\,,7\,,\\&&\\
\displaystyle{(1-u)^\frac{1}{3} g(u) \delta^a_\mu }& \qquad & a=2\,,3\,,4\,
\end{array}
\right.
\]
and
\[
\displaystyle{\left.\frac{\partial^2 s^a}{\partial t^\mu \partial t^{\mu^\star}} \right|_{\bf 0^\star}}= \left\{
\begin{array}{lcl}
\frac{1}{9} s^2 (1-u) g(u) g^\prime(u) & \qquad & a=1\,,\\ &&\\
0 & & a\neq 1\,.
\end{array}
\right.
\]
Thus
\[
\sum_{\mu=1}^8 c_{\mu\mu^\star 2,7}(t) =
8 \left.\frac{\partial s^1}{\partial t^2 \partial t^7}\right|_{\bf 0^\star} +
\left.
\frac{\partial s^2}{\partial t^2}\right|_{\bf 0^\star}.
\left.
\frac{\partial s^7}{\partial t^7}\right|_{\bf 0^\star}.
\sum_{p=1}^8 c^p_{p2,7}( {\bf s} )\,.
\]
Note, it is important {\sl not} to lower the $p$-index in the final sum in the above equation. This is because the ${\bf s}$-variables
are not flat-coordinates and hence the metric that one would use to lower an index will not have constant entries.

\medskip

To calculate $c^p_{p2,7}(s)$ (which, again, is a dimensional quantity and hence a function of $s$ alone) one just requires
those terms linear in the variable $s_7\,.$ Thus it suffices to calculate the structure functions using the reduced
superpotential
\[
W^{reduced} = x^3+y^3+z^3 + s\, x y z + s_7\, xy\,.
\]
Using as a basis for the ring $\phi_1=1\,,\phi_2=z\,,\phi_3=y\,,\phi_4=x\,,\phi_5=yz\,,\phi_6=xz\,,\phi_7=xy\,,\phi_8=xyz$
(so the multiplication in the ring is $\phi_a.\phi_b=c_{ab}^c({\bf s}) \phi_c$) it is straightforward to show that
\[
c^1_{12}=c^2_{22}=c^3_{32}=c^4_{42}=c^7_{72}=0
\]
and
\[
c^5_{52}=c^6_{62}=c^8_{82} = -\frac{2}{27} \frac{s^2}{(1-u)} s_7\,.
\]
Hence
\[
\sum_{p=1}^8 c^p_{p2,7}(s) = -\frac{2}{9} \frac{s^2}{1-u}\,.
\]
With this one obtains
\[
\sum_\nu c_{\nu\nu^\star 2,7}(t) =\frac{8}{9} s^2 (1-u) g(u) g^\prime(u) - \frac{2}{9} g(u)^2 s^2\,.
\]

\medskip

To convert this expression into a function of $t$ one requires certain properties of the hypergeometric equation.
For example, $g(u)$ and $u^\frac{1}{3} {}_2F_1(\frac{2}{3},\frac{2}{3},\frac{4}{3},u)$ are linearly independent solutions
of the hypergeometric equation with constants $a=\frac{1}{3}\,,b=\frac{1}{3}\,,c=\frac{2}{3}\,.$ Hence, on using properties
of the Wronskian for the hypergeometric equation one can easily show that
\[
\frac{dt}{ds}= \frac{1}{(1-u)} \frac{1}{g(u)^2}
\]
and hence that
\[
s^2 g(u)^2 = 9 \frac{d~}{dt} \log(1-u)
\]
and
\[
s^2 (1-u) g(u)g^\prime(u) = - \frac{9}{2} \frac{s^{\prime\prime}(t)}{s^\prime(t)} + \frac{9}{2} \frac{d~}{dt} \log(1-u)\,.
\]
With these, equation (\ref{GE6}) becomes
\[
\frac{dG}{dt} = \frac{1}{48} \left[ 2 \frac{d~}{dt} \log(1-u) - 4 \frac{d~}{dt} \log s^\prime(t) \right]
\]
and so finally (and recall $G$ is only defined up to a constant),
\[
G(t)=\displaystyle{-\frac{1}{24} \log \left\{ \frac{    \left[ s(t)^\prime \right]^2 }{1+\frac{1}{27} s(t)^3 }\right\}}\,.
\]

\bigskip

The calculations for $\widetilde{E}_{7,8}$ follow in exactly the same way. The only main difference is that certain of
the $s^a$-variables are linear in {\sl two} of the $t^\mu$-variables, so extra terms appear. Also, for a purely computational
perspective, it is easier to calculate (say, for $\widetilde{E}_8$) $c^\nu_{\nu 2,8}$ rather than $c^\nu_{\nu 8,2}\,.$ These two
expressions must be, of course, identical, but to prove this requires the use of some non-trivial quadratic hypergeometric function identities
\cite{Matsumoto}.

\endproof

\medskip

\noindent These $G$-functions have simple transformation properties under the modular group:

\begin{lemma}
For $\widetilde{E}_{6,7,8}$ (so $n=8\,,9\,,10$):
\begin{eqnarray*}
G(t+1) & = & G(t)\,,\\
G\left(-\frac{1}{t}\right) & = & G(t) + \left(\frac{n}{24}- \frac{1}{2} \right) \log t\,.
\end{eqnarray*}
\end{lemma}
{\noindent{\bf Proof~} This result follows immediately from the fact that the functions $s=s(t)$ are modular invariant.
\endproof

Note that the flat coordinates \cite{NY} are defined in a disc, and that these modular transformations take one outside the original domains of
definition. However one may glue together these objects to obtain a so-called twisted Frobenius manifold \cite{D1}. The origin of such
modularity is the existence of a hidden symmetry in the underlying Frobenius manifold. A symmetry of the WDVV equations is a transformation
\begin{eqnarray*}
t^\alpha & \mapsto & {\hat{t}^\alpha}\,, \\
g_{\alpha\beta} & \mapsto & {\hat g}_{\alpha\beta}\,,\\
F & \mapsto &{\hat F}
\end{eqnarray*}
that acts on the solution space of the WDVV equations. One such symmetry, denoted $I$ in \cite{D1}, is the following:
\begin{eqnarray*}
{\hat t}^1 & = & \frac{1}{2} \frac{t_\sigma t^\sigma}{t^{n}}\,,\\
{\hat t}^i & = & \frac{t^i}{t^{n}}\,, \quad i = 2\,,\ldots\,,n-1\,,\\
{\hat t}^{n} & = & - \frac{1}{t^{n}}\,,\\
&&\\
{\hat \eta}_{\alpha\beta} & = & {\eta}_{\alpha\beta}\,,\\
&&\\
{\hat F}({\hat{\bf t}})&  = &\left(t^{n}\right)^{-2} \left[ F({\bf t}) - \frac{1}{2} t^1 (t_\sigma t^\sigma) \right]
\end{eqnarray*}
which induces a symmetry of the WDVV equations. Up to a simple equivalence, $I^2 = \mathbb{I}\,.$
Under such a transformation (which is, in fact, a Schlesinger tranformation) $G$ transforms as
\[
{\hat{G}}\left({\hat{{\bf t}}}\right)  =  G({\bf t}) + \left(\frac{n}{24}- \frac{1}{2} \right) \log t^n\,.
\]
Generically $I$ is a symmetry between two (apparently) dissimilar manifolds. However for the simple elliptic
symmetries (and more generally any Frobenius manifold with $d=1$ and $d^n=0$) the corresponding manifolds sit at the
fixed point of the involutive symmetry and hence one obtain a hidden symmetry of the manifold under the action of
the modular group.

\medskip

The $G$-function is a function of the marginal variable through certain very special functions, each of which
determines certain singular loci within the Frobenius manifold. It is straightforward, using the explicit formulae in \cite{NY}, to show that the Jacobian
determinant of the transformation between the $\{{\bf s}\}$ and $\{{\bf t}\}$ coordinates is
\[
\left|
\frac{\partial (s^1\,,\ldots\,,s^8)}{\partial (t^1\,,\ldots\,,t^8)}
\right| = \left( \frac{ds}{dt}\right)^4\,.
\]
With this the $G$-function for $\widetilde{E}_6$ can be written as
\[
G(t)=-\frac{1}{24} \log \left\{ \frac{1}{1+\frac{1}{27} s^3} \,. \left\vert \frac{\partial (s^1\,,\ldots\,,s^8)}{\partial (t^1\,,\ldots\,,t^8)} \right\vert^\frac{1}{2} \right\}\,.
\]
Thus the building blocks of $G$ are structures the vanishing of which determines certain singular structures:
\begin{itemize}
\item[$\bullet$] if $1+\frac{1}{27}s^3=0$ then the ring multiplication breaks down;
\item[$\bullet$] if $\displaystyle{\left| \frac{\partial (s^1\,,\ldots\,,s^8)}{\partial (t^1\,,\ldots\,,t^8)} \right\vert}=0$ then the system of flat coordinates breaks down
\end{itemize}
(and similar results hold for $\widetilde{E}_{7,8}$). Thus the $G$-function contains information not only the
variance of the spectral data, but also on the singular, or boundary, structures within the Frobenius manifold.
It would be of interest to understand the nature of the $F$-manifold structure on such degenerate objects.

The isomonodromic $\tau$-functions for the Coxeter group orbits spaces has a very simple form \cite{KS}:
\[
\tau_I= \prod_{i=1}^{\#\K_i} \kappa_i^{-\frac{(N_i-2)^2}{16 N_i}}\,,
\]
where the components of the caustic are given in terms of quasi-homogeneous irreducible polynomials $\kappa_i$
such that $\kappa_i^{-1}(0)=\K_i\,.$ In particular\footnote{The Dubrovin and Zhang $\tau_I$ differs from
the usual one by a factor of $-2$, so the conventional form would read $\tau_I=\kappa_1^{\frac{1}{24}}\,.$},
for the simple hypersurface singularities which have only one component
to the caustic (see Table 1)
\[
\tau_I^{-48}=\kappa_1\,.
\]
Using properties of multidimensional resultants one can show, for $\widetilde{E}_6$ and similarly for $\widetilde{E}_{7,8}\,,$ that
\[
\tau_I^{-48} =
\frac{    \left[ s(t)^\prime \right]^4 }{\left[1+\frac{1}{27} s(t)^3\right]^2 } \det\left\{
m_{Hess}: \mathcal{R}\rightarrow \mathcal{R} \right\}\,
\]
where $m_{Hess}$ denotes multiplication by the Hessian in the Jacobi ring $\mathcal{R}\,.$

\section{Codimension-one elliptic root systems}

The Frobenius manifolds arising from the singularities $\widetilde{E}_{6,7,8}$ belong to another family of manifolds constructed from
elliptic root systems. Such families were introduced by K. Saito \cite{saito3} and studied further by Satake \cite{SatakeFrob} . In this
section the $G$-functions will be constructed for low-dimensional examples of such manifolds where the prepotential is known explicitly.
This is achieved by direct calculations (using Mathematica) using Theorem \ref{virasoro} to first calculate the first derivatives of the $G$-function.

\begin{example}

The prepotential for the $D_4^{(1,1)}$ orbit space has been directly calculated by Satake \cite{SatakeD4} . Explicitly\footnote{In this example  $t_i$=$t^i$ for notational convenience} (and after a few redefinitions):
\begin{eqnarray*}
F & = & \phantom{-} \frac{1}{4} t_1^2 t_6+ \frac{1}{2} t^1 \left( t_2^2 + t_3^2 + t_4^2 + t_5^2 \right)  - \frac{1}{24} \left( t_2^4 + t_3^4 + t_4^4 + t_5^4 \right) (u + v + w )  \\
&&-\frac{1}{4} \left( t_2^2 t_3^2 + t_2^2 t_4^2 + t_2^2 t_5^2 + t_2^3 t_4^2 + t_3^2 t_5^2 +  t_4^2 t_5^2 \right)
(u + v ) - t_2 t_3 t_4 t_5 ( u - v)
\end{eqnarray*}
where the functions $u\,,v\,,w$ depend on the variable $t_6$ alone and satisfy the Halphen system
\begin{eqnarray*}
\dot{u} & = & u\,v + u\,w - v\,w\,,\\
\dot{v} & = & v\,u + v\,w - u\,w\,,\\
\dot{w} & = & w\,u + w\,v - u\,v\,.
\end{eqnarray*}
Solutions of this system have been discussed by a number of authors (see, for example, \cite{D1} Appendix C). The Euler vector field takes the form
\[
E=t_1 \frac{\partial}{\partial t_1} + \frac{1}{2} \left(\frac{\partial}{\partial t_2}+\frac{\partial}{\partial t_3}+\frac{\partial}{\partial t_4}+\frac{\partial}{\partial t_5}\right)
\]
and have $d=1\,.$ It is easy to show that $\gamma=0\,.$ The $G$-function for this manifold takes the form
\[
G(t_6) = -\frac{1}{2} \log \eta(t_6)
\]
where $\eta$ is the Dedekind $\eta$-function.

\end{example}

Other examples of prepotentials may be obtained from this example by the process of folding. The notation $B\hookrightarrow A$ will be used
to denote the process of obtaining the manifold $B$ from $A$ by such a method (so, for example, $F_4 \hookrightarrow E_6$ in the Coxeter case).
The examples $B_3^{(1,1)}\,, B_2^{(2,1)}$ and $G_2^{(1,1)}$ may all be obtained from $D_4^{(1,1)}$ by similar foldings to planar submanifolds $\Sigma\,.$

\begin{example}
For these three Frobenius manifolds, obtained by folding the above example, the $G$-function takes the universal form
\[
G=\left(\frac{\gamma}{{\rm deg}(\kappa)}\right) \log \kappa - \frac{1}{2} \log\eta(t^n)\,,
\]
where $\gamma$ is the scaling anomaly, ${\rm deg}\kappa = E(\log \kappa)$ and $\kappa^{-1}(0)$ is a caustic submanifold where the (generically semi-simple) multiplication becomes nilpotent. The submanifold $\Sigma$ and other data is given in Table 2.

\begin{table}
\begin{center}
\begin{tabular}{c|c|c|c}
Elliptic root system & $\gamma$ & ${\rm deg}{(\kappa)}$ & submanifold $\Sigma$ \\ \hline
$B_3^{(1,1)}$ & $-\frac{1}{48}$ & $1$ & $\{ t_4=t_5 \}$ \\
&&&\\
$B_2^{(2,1)}$ & $-\frac{1}{24}$ & $2$ & $\{ t_2=t_3\,,t_4=t_5 \}$ \\
&&&\\
$G_2^{(1,1)}$ & $-\frac{1}{24}$ & $\frac{1}{2}$ & $\{t_3=t_4=t_5 \}$
\end{tabular}
\end{center}
\vskip 5mm
\caption{Data for the foldings of $D_4^{(1,1)}$}
\end{table}

\noindent The final folding to a three dimensional manifold results in the $A_1^{(1,1)}$-example (see next section).

\end{example}

The $\widetilde{E}_{6,7,8}$ singularities have embedded within them various unimodal boundary singularities. Thus, using
Arnold's notation,
\begin{eqnarray*}
L_6 = D_{4,1} & \hookrightarrow & P_8=\widetilde{E}_{6}\,,\\
K_{4,2} & \hookrightarrow & X_9=\widetilde{E}_{7}\,,\\
F_{1,0} & \hookrightarrow & J_{10}=\widetilde{E}_{8}\,.
\end{eqnarray*}
In terms of the function $W$ on has
\[
W(z_1,z_2\,,\ldots\,,z_n) = \overline{W}(z_1^2,z_2\,,\ldots\,,z_n)
\]
so the boundary is given $z_1=0\,.$ In principle, therefore, one may obtain the corresponding prepotential by performing a similar
restriction and hence obtain the $G$-function for these unimodal boundary singularities by direct calculation. This, though, is beyond
the capabilities of Mathematica (and would not shred much light onto the structure of these manifolds), and without knowledge of the
$F$-manifold structure of these manifolds the techniques developed earlier cannot be used.

However, a simpler reduction exists in the ${\widetilde{E}}_6$ case which utilizes the $\mathbb{Z}_3$-symmetry in this model. The $W$
function is given by the restriction to the submanifold $\{s_7=s_6=s_5\,,s_4=s_3=s_2 \}$, i.e.
\[
W({\bf z},{\bf s}) = z_1^3+z_2^3+z_3^3 + s_8 \, z_1 z_2 z_3 + s_6 \, (z_1 z_2 + z_2 z_3 + z_3 z_1) + s_3 \, (z_1+z_2+z_4) + s_1\,.
\]
Using Noumi's expressions for the flat coordinates one obtains the submanifold
\[
\Sigma = \{ t^7=t^6=t^5 \,, t^4=t^3=t^2 \}\,,
\]
and in terms of elliptic root systems this corresponds to the folding $G_2^{(3,1)} \hookrightarrow E_6^{(1,1)}\,.$
Direct computations, again with Mathematica, gives
\[
G=-\frac{1}{12} \log \kappa_1 + \frac{5}{24} \log \lambda_t\,, \qquad\qquad \gamma=-\frac{1}{18}\,,
\]
where $\kappa_1=\alpha(t^8)\, t^3 + \beta(t^8) (t^6)^2\,,$ for some functions $\alpha$ and $\beta\,.$ In this
four-dimensional case the isomonodromic $\tau$-function can also be calculated,
\[
\tau_I^{-48} = f(t^8) \, \kappa_1^8 \, \kappa_2
\]
where $\kappa_1^{-1}(0)\,, \kappa_2^{-1}(0)$ are irreducible caustics and $f$ a function of the marginal variable alone. Comparing this with Theorem \ref{ClausThm}
suggests that in this case there are two caustics with $N_1=6$ and $N_2=3\,.$ Similar conjectures may also be made for
the other examples constructed in this section.

Note that in all these cases the functions $G$ and $\tau_I$ are far simpler in form than one might expect from their very complicated
definitions - this suggests some deeper structure behind these functions. Note also that while, for boundary singularities
\[
{\overline{F}} = F|_\Sigma\,,
\]
the corresponding $G$-functions do not restrict in the same way. Further explicit knowledge of the $F$-manifold structures of boundary singularities is clearly required.

\section{Discussion}

One of the main motivations in constructing the $G$-function for these singularities is to construct a corresponding
dispersive integrable hierarchy. Dispersionless systems, examples of equations of hydrodynamic type
\[
u^i_T = A^i_j({\bf u}) u^j_X\,,
\]
may be constructed from the known prepotentials of $\widetilde{E}_{6,7}\,;$ this is entirely algorithmic. Even for
$\widetilde{E}_8$ general properties of the hierarchy may be derived \cite{NY}. For example,  the action of the symmetry $I$ lifts from the
manifold to the flows, resulting in these flows having a hidden modular symmetry \cite{MS}. The $G$-function appears in the construction
of first order deformations of these flows, i.e. in the construction of first order dispersive terms. Thus first order deformations of the equations of hydrodynamic type associated with these singularities can be written down using the results of Dubrovin and Zhang \cite{DZ2}.

Similar results have also been obtained for the Jacobi group orbit spaces $\Omega/J(A_N)\,$ (which are of dimension $n=N+2$). The
explicit form of the $G$-function\footnote{The notation here is different from the rest of this paper: the reader should consult \cite{KS} for the precise definitions of the variables which appear in this formula.} was conjectured by the author in \cite{me}  and proved in \cite{KK} (see also \cite{KS}),
\[
G=-\log\eta(t_0) - \frac{n+2}{24} \log t_{n+2}
\]
where $\eta$ is the Dedekind $\eta$-function.
This construction utilized an alternative description of this orbit space as the Hurwitz space $H_{1,N}(N)$, the moduli space of $N$-fold branched  coverings $p: \mathbb{T}\rightarrow \mathbb{P}^1$ where $\mathbb{T}$ is the (genus 1) torus. These examples also have $d=1\,,$ $d^n=0$ and hence
sit at the fixed point of the symmetry $I\,.$  Again, the
dispersionless integrable systems may be written down and their first order deformation constructed.

The ultimate goal of this work is to construct a fully dispersive integrable system associated to the simple elliptic singularities
(and more generally, the Jacobi group orbit spaces \cite{B}).
Central to this programme is the construction and properties of the total descendent potential \cite{GM}
\[\DD^M = e^{G}\cdot \widehat S_\tau^{-1}\cdot \widehat \Psi\cdot
(\exp^{U/z})^{\widehat{}} \left(\prod^n_{i=1}\DD^{A_1}\right)\,.
\]
Here $S_\tau$, $\Psi$ and $\exp^{U/z}$
are certain objects related to the underlying isomonodromy problem
associated to any (semi-simple) Frobenius manifold and hats indicate quantizations of the classical data. Note that $G$-appears explicitly in such formulae.
For the simple singularities this programme has been realized:  Givental
and Milanov \cite{GM} showed how this total descendent potential can be constructed from properties of these singularities and also how a vertex description of the associated hierarchies can be constructed. As remarked in their paper, the unimodular simple elliptic singularities (together with the Jacobi group orbits spaces) are the next cases that should be considered in this programme,
and new results have recently been obtained \cite{KrSh,MR}. The elliptic singularities base spaces also turn out to be mirror to certain  orbifold quantum cohomology manifolds \cite{ST}. The results obtained here show that the $G$-function extends holomorphically over the caustic for these more complicated examples. How the hidden symmetry $I$ acts on such dispersive hierarchies is an interesting question: one expects the appearance of modular invariant dispersive hierarchies.

\bigskip

\noindent{\bf Acknowledgment:} I would like to acknowledge financial support from the British Council (PMI2 Research Cooperation award).
I would also like to thank Claus Hertling for many conversations on caustics, $F$-manifolds and $G$-functions over the past few years and
Claire Gilson for pointing out \cite{Matsumoto} to me.

\end{document}